\documentclass[twoside,fleqn]{article}
\usepackage{espcrc2}
\usepackage{graphicx}

\title{
{\Large  The $\tau^+\tau^-$ production cross section near threshold revisited}}
\author{P. Ruiz-Femen\'\i a
\address{
       {\em Instituto de F\'\i sica Corpuscular, Universitat de Val\`encia,} \\
       {\em Apartat Correus 22085, E-46071 Val\`encia, Spain}}
       } 

\begin{document}

\newcommand{\nn}{\nonumber}
\newcommand{\mev}{\mbox{\rm MeV}}
\newcommand{\gev}{\mbox{\rm GeV}}
\newcommand{\eqn}[1]{(\ref{#1})}
\newcommand{\MSb}{{\overline{MS}}}
\newcommand{\ep}{\epsilon}
\newcommand{\IM}{\mbox{\rm Im}}
\newcommand{\lsim}{\stackrel{<}{_\sim}}
\newcommand{\gsim}{\stackrel{>}{_\sim}}

\begin{abstract}
Next-to-next-to-leading contributions to the
cross section $\sigma(e^+e^-\to\tau^+\tau^-)$ at energies
close to threshold are analysed,
taking into account the known 
non-relativistic effects
and ${\cal{O}}(\alpha^2)$ corrections. The numerical changes
with respect to previous works are
small, but the new corrections give a true estimate 
of the uncertainty in the theoretical calculation.  
\end{abstract}

\maketitle

\newcommand{\jhep}[3]{{\it JHEP }{\bf #1} (#2) #3}
\newcommand{\nc}[3]{{\it Nuovo Cim. }{\bf #1} (#2) #3}
\newcommand{\npb}[3]{{\it Nucl. Phys. }{\bf B #1} (#2) #3}
\newcommand{\npps}[3]{{\it Nucl. Phys. }{\bf #1} {\it(Proc. Suppl.)} (#2) #3}
\newcommand{\plb}[3]{{\it Phys. Lett. }{\bf B #1} (#2) #3}
\newcommand{\pr}[3]{{\it Phys. Rev. }{\bf #1} (#2) #3}
\newcommand{\prd}[3]{{\it Phys. Rev. }{\bf D #1} (#2) #3}
\newcommand{\prl}[3]{{\it Phys. Rev. Lett. }{\bf #1} (#2) #3}
\newcommand{\prep}[3]{{\it Phys. Rep. }{\bf #1} (#2) #3}
\newcommand{\zpc}[3]{{\it Z. Physik }{\bf C #1} (#2) #3}
\newcommand{\sjnp}[3]{{\it Sov. J. Nucl. Phys. }{\bf #1} (#2) #3}
\newcommand{\jetp}[3]{{\it Sov. Phys. JETP }{\bf #1} (#2) #3}
\newcommand{\jetpl}[3]{{\it JETP Lett. }{\bf #1} (#2) #3}
\newcommand{\ijmpa}[3]{{\it Int. J. Mod. Phys. }{\bf A #1} (#2) #3}
\newcommand{\hepph}[1]{{\tt hep-ph/#1}} 
\newcommand{\hepth}[1]{{\tt hep-th/#1}} 
\newcommand{\heplat}[1]{{\tt hep-lat/#1}}



\section{Introduction}

The Tau--Charm Factory, a high--luminosity ($\sim 10^{33}\;\mbox{cm}^{-2}\;
\mbox{s}^{-1}$) $e^+e^-$ collider with a centre--of--mass energy
near the $\tau^+\tau^-$ production threshold, has been proposed
as a powerful tool to perform high--precision studies
of the $\tau$ lepton, charm hadrons and the charmonium system
\cite{marbella}.
In recent years, this energy region has been only partially explored
by the Chinese BEBC machine ($\sim 10^{31}\;\mbox{cm}^{-2}\;\mbox{s}^{-1}$).
The possibility to operate the Cornell CESR collider
around the $\tau^+\tau^-$ threshold \cite{cornell}
has revived again the interest on Tau--Charm Factory physics. Also
the CMD-2 detector at the Novosibirsk VEPP-2M $e^+e^-$ collider 
will be soon ready to collect new data in this region 
\cite{eidelman}. 

A precise understanding of the $e^+e^-\to\tau^+\tau^-$ production
cross section near threshold is clearly required. The accurate
experimental analysis of this observable could allow to improve the
present measurement \cite{BES} of the $\tau$ lepton mass.
The cross section $\sigma(e^+e^-\to\tau^+\tau^-)$ has already been 
analysed to ${\cal O} (\alpha^3)$ in refs.~\cite{voloshin},
including a resummation of the leading Coulomb corrections.

The recent development of non-relativistic effective
field theories of QED (NRQED) and QCD (NRQCD)  \cite{lepage}
has allowed an extensive investigation of the threshold production of
heavy flavours at $e^+e^-$ colliders. The threshold $b\bar b$ 
\cite{jamin}
and $t\bar t$ \cite{topprodsummary}
production cross sections have been computed to the
next-to-next-to-leading order (NNLO) in a combined expansion in powers of
$\alpha_s$ and the fermion velocities.
Making appropriate changes, those calculations can be easily applied to
the study of $\tau^+\tau^-$ production \cite{ruiz}. One can then achieve
a theoretical precision better than 0.1\%.

\section{Perturbative calculation to ${\cal{O}} (\alpha^4)$}
\label{sec:pert}

At lowest order in QED, the $\tau$ leptons are produced by one-photon exchange
in the s-channel, and the total cross section formula reads
\begin{equation}
\sigma_{\mbox{\tiny $B$}} (e^+\,e^- \to \tau^+ \, \tau^-) ={{2 \pi \, \alpha^2} \over {3s}}\,
v\, (3-v^2)\,,
\label{s0}
\end{equation}
where $v=\sqrt{1-4 m_{\tau}^2/s}$ is the velocity of the final $\tau$ 
leptons in the center-of-mass 
frame of the $e^+ \, e^-$ pair which makes $\sigma_B$ vanish when $v\to 0$. 

Electromagnetic corrections of ${\cal O} (\alpha)$ arise from the
interference between the tree level result and 1-loop amplitudes. 
A factor $\alpha/v$ emerges in the 1-loop final
state interaction between the tau leptons, making the cross section at threshold
finite. Furry's theorem guarantees that contributions to 
$\sigma(e^+e^-\to \tau^+\tau^-)$ coming from initial, intermediate and final
state corrections completely factorize at ${\cal O} (\alpha^3)$, including real
photon emission.

\begin{figure}[tb!]
\vspace*{0.4cm}
\hspace*{-0.3cm}
\includegraphics[angle=0,height=2.0cm,width=0.5\textwidth]{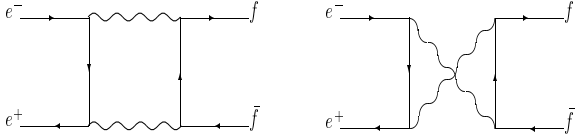}
\vspace*{-1.0cm}
\caption[]{\label{fig:boxes} \it Direct and crossed  
box diagrams for $e^+e^-\to f\bar{f}$.}
\vspace*{-0.5cm}
\end{figure}

Some undesirable features appear at ${\cal O} (\alpha^4)$: The two-loop
$\tau^+\tau^-\gamma$ vertex develops an $\alpha^2/v^2$ term which makes 
the cross section ill-defined when $v\to0$, and multiple photon production
of tau leptons by box-type diagrams and the non-zero interference of initial and final state radiation
spoil exact factorization. However, as it has been recently shown \cite{box},
the squared amplitude of the $e^+e^-\to \tau^+\tau^-$ box diagram 
(see Fig.~\ref{fig:boxes}) is
proportional to $\alpha^4v^2$, and so represents a N$^4$LO correction in the
combined expansion in powers of $\alpha$ and $v$, far beyond the scope of this
analysis. In addition, contributions to the total cross section from
diagrams with real photons emitted from the produced taus can be shown to
begin at N$^3$LO, and factorization remains 
at NNLO. The total cross section can thus be 
written as an integration over
the product of separate pieces including initial, intermediate and
final state corrections:
\begin{equation}
\sigma(s)=\int^{s} \, F(s,w)\,\bigg|\frac{1}{1+e^2\Pi{\mbox{\tiny em}}(w)}
\bigg|^{2}\,
\tilde{\sigma}(w)\,dw \,.
\label{secradR}
\end{equation}
The radiation function $F(s,w)$ \cite{kuraev} describes initial state radiation,
including virtual corrections. 
The integration accounts for the effective energy loss
due to photon emission from the $e^+ \, e^-$ pair, and it
includes the largest corrections coming from the
emission of an arbitrary number of initial photons, which can sizable
suppress the total cross section. $\tilde{\sigma}$ collects only
final-state interactions between the tau leptons, and it is 
usually written in terms of the tau spectral density 
$R_{\mbox{\tiny $\tau$}}$,
\begin{equation}
\tilde{\sigma}(e^+e^-\to\gamma^*\to \tau^+\tau^-)=
R_{\mbox{\tiny $\tau$}}(s)  
\,\sigma_{pt}\,,
\label{R(s)tau}
\end{equation}
with $\sigma_{pt}\, = \,
\frac{4\pi\alpha^2}{3s}$.
The threshold behaviour of the
total cross section will be ruled by the expansion of
$R_{\mbox{\tiny $\tau$}}$ at low velocities.
  
 
\section{Non-Relativistic Corrections: NRQED}
\label{sec:NRQED}

The aim of the effective field theory approach to threshold 
particle production is to achieve a given accuracy in the
combined expansion in powers of $\alpha$ and the velocity $v$.
Such double expansion is needed at threshold due
to the appearance of $(\alpha/v)^n$ terms in the QED perturbative
expansion at the $n$-th loop order, which forces one to
treat $\alpha$ and $v$ on the same footing. A non-perturbative
procedure to deal with such singular terms in the limit
$v\to 0$ is therefore mandatory. 

\begin{figure}[tb!]
\hspace*{1.0cm}
\includegraphics[angle=0,height=3.0cm,width=0.3\textwidth]{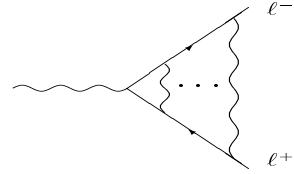}
\vspace*{-1.0cm}
\caption[]{\label{fig:coulomb} \it Ladder exchange of photons between
produced fermions.}
\vspace*{-0.5cm}
\end{figure}

The leading divergences (i.e.
$\left( \frac{\alpha}{v} \right)^n$, $n > 1$), emerging
from the ladder diagrams with Coulomb photons shown in Fig.~\ref{fig:coulomb},
are resummed in the 
well-known Sommerfeld factor \cite{sommer}
\begin{equation}
|\Psi_{c,\mbox{\tiny E}}(0)|^2 \, = \, \frac {\alpha\pi/v}{1-\mbox{exp}(-
 \alpha\pi/v)}\,,
\label{culfactor}
\end{equation}
multiplying the Born cross section~(\ref{s0}).
Next-to-leading order (NLO) entails terms proportional to
$(\alpha/v)^n\times [\alpha,v]$, while NNLO accuracy stands
for contributions $(\alpha/v)^n\times [\alpha^2,v^2,\alpha\,v]$.

A systematic
way to calculate these higher-order corrections 
in this regime requires the use of a simplified
theory which keeps the relevant physics at the scale $Mv \sim M\alpha$,
characteristic of the Coulomb interaction.
NRQED \cite{lepage} was designed precisely for this purpose. It is an effective
field theory of QED at low energies, applicable to fermions in non-relativistic
regimes, i.e. with typical momenta $p/M \sim v \ll 1$. Interactions contained in the
NRQED Lagrangian have a definite velocity counting but
propagators and loop integrations can also generate powers of $v$. With appropriate
counting rules at hand, one can prove that all interactions between the non-relativistic
pair $\tau^+\tau^-$ can be described up to NNLO in terms of time-independent potentials
\cite{labelle}, derived from the low-energy Lagrangian. Therefore,
the low-energy expression of the $\tau$ spectral density is related with the
non-relativistic Green's functions \cite{hoangteubner}:
\begin{equation}
R^{\mbox{\tiny NNLO}}_{\mbox{\tiny $\tau$}}(s)  = 
\frac{6\,\pi}{M^2} \,\, \mbox{Im} \Big( C_1\,
G(E)
\,-\frac{4E}{3M} \,
G_c(E)\Big)
\label{R(s)NNLOmain}
\end{equation} 
with $C_1$ a short distance coefficient to be determined by matching full and effective
theory results and $E=m_{\tau}v^2$ the non-relativistic energy.
The details of this derivation can be found
in the Appendix B of \cite{ruiz}.
The Green's function $G$ obeys the Schr\"odinger equation corresponding 
to a two-body system
interacting through potentials derived from ${\cal{L}}_{\mbox{\tiny NRQED}}$
at NNLO:
\begin{eqnarray}
\bigg(\!\! -\frac{{\mbox{\boldmath $\nabla$}}^2}{M} 
\!\!\!\!\!&-&\!\!\!\!\!
\frac{{\mbox{\boldmath $\nabla$}}^4}{4M^3} +
V_{c}({\mbox{\boldmath $r$}}) + V_{\mbox{\tiny BF}}({\mbox{\boldmath $r$}})
+ V_{\mbox{\tiny An}}({\mbox{\boldmath $r$}})
-E\bigg)\nonumber\\
&\times& \!\!\!\!G({\mbox{\boldmath $r$}},{\mbox{\boldmath $r$}^\prime},E)
\, = \,
\delta^{(3)}({\mbox{\boldmath $r$}}-{\mbox{\boldmath $r$}^\prime})
\label{Schrodingerfull}
\end{eqnarray}
The term  $-\frac{{\mbox{\boldmath $\nabla$}}^4}{4M^3}$
is the first relativistic correction to the
kinetic energy. $V_c$ stands for the Coulomb potential with 
${\cal O}(\alpha^2)$ corrections. 
At NNLO, the heavy leptons are only produced in triplet 
S-wave states, so we just need to consider the corresponding projection of the 
Breit-Fermi potential $V_{\mbox{\tiny BF}}$.
Finally, $V_{\mbox{\tiny An}}$ is a NNLO piece
derived from a contact
term in ${\cal{L}}_{\mbox{\tiny NRQED}}$, 
which reproduces the QED tree level
s-channel diagram for the process $\tau^+\tau^-\to \tau^+\tau^-$.

A solution of eq.~(\ref{Schrodingerfull}) must rely on numerical or perturbative
techniques. In the QED case, a significant difference between both approaches is
not expected, being $\alpha$ such a small parameter.
Consequently we follow the perturbative approach, using recent
results \cite{hoangteubner}, 
where 
NLO and NNLO corrections to the Green's
function are calculated analytically, via
Rayleigh-Schr\"odinger time-independent perturbation theory around the
known LO Coulomb Green's function $G_c$. We refer the reader to Appendix C
of \cite{ruiz} for complete expressions of the Green's function corrections.

\section{Vacuum Polarization}
\label{sec:vp}

For a complete NNLO description of $\sigma(e^+e^- \to \tau^+\tau^-)$,
two-loop corrections to the photon propagator should be included.
The light lepton
contributions to the vacuum polarization are the standard
1- and 2-loop perturbative expressions.
For the $\tau$ contribution in the threshold vicinity $q^2 \gsim 4M^2$, 
resummation
of singular terms in the limit $v\to 0$ is again mandatory. 
Under the assumption
$\alpha \sim v$, we need to know NLO contributions to
$\Pi_{\tau}(q^2)$, performing the direct
matching for both real and imaginary parts.

In the hadronic sector, we can relate
the hadronic vacuum polarization with the total cross section
$\sigma(e^+e^-\to\gamma^*\to had)$.
Below 1~GeV, the electromagnetic
production of hadrons is dominated by the $\rho$ resonance
and its decay to two charged pions. The photon mediated $\pi^+\pi^-$
production cross section is
driven by the pion electromagnetic form factor $F(s)$.
An analytic expression for $F(s)$
was obtained in Ref.~\cite{guerrero}, using Resonance
Chiral Theory and the restrictions
imposed by analyticity and unitarity. The obtained $F(s)$
provides an excellent description of experimental data up to energies
of the order
$s_{\rho} \sim $ 1~GeV$^2$.

For the integration region above
$s_{\rho}$, we use $\mbox{Im}
 \Pi_{\mbox{\tiny had}}$ as calculated from pQCD.
Our simple estimate has been proved \cite{ruiz} to deviate
by less than 5$\%$ for the running of $\alpha$ at the scale
$\sqrt{s}=M_Z$. Considering
that $\Pi_{\mbox{\tiny had}}$ modifies
$\sigma(e^+e^-\to \tau^+\tau^-)$ near threshold by roughly $1\%$, 
our result has a global
uncertainty smaller than $0.1\%$ for the total cross section. Clearly,
our antsaz could be easily improved using a more realistic hadronic
spectrum, and, indeed, most experiments have their own
routines to accurately implement 
vacuum polarization together with initial state radiation 
in their data analyses. 


\section{Electroweak and Bound States effect}
\label{sec:ew}

Electroweak production of heavy quarks
including threshold effects
has already been studied in previous papers \cite{hoang-teubner2},
and can be easily incorporated in our
basic formula (\ref{secradR}). However, the characteristic $m_{\tau}^2/M_Z^2$
suppression of this set of corrections and the small value of the
neutral-current couplings of the leptons make these terms fully negligible in
our analysis. 

Green's functions develop energy poles below threshold corresponding
to spin triplet (n$^3S_1$) electromagnetic $\tau^+\tau^-$ bound states.
The small width 
of these bound levels, 
dominated by their $e^+e^-$ decay rate
\begin{equation}
\Gamma_{ee}=\frac{m_{\tau}\alpha^5}{6n^3}\approx 
\frac{6.1\cdot 10^{-3}\,\,\mbox{eV}}{n^3}\,,
\nonumber
\end{equation}
make these states very difficult to be resolved experimentally.
For the same reason, the bound states cannot affect the shape of
the cross section at threshold, contrary to the case of heavy quark threshold
production, where bound states play a crucial role.


\section{Numerical analysis for $\sigma(e^+e^- \to \tau^+\tau^-)$}
\label{sec:numerics}

\begin{figure}[!tb]
\begin{center}
\hspace*{-0.5cm}
\includegraphics[angle=0,height=5.5cm,width=0.5\textwidth]{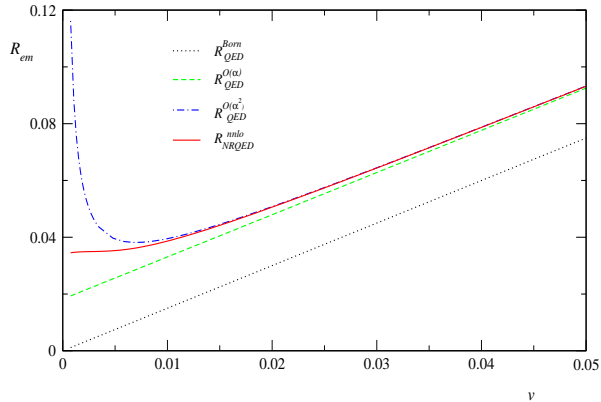}
\end{center}
\vspace*{-1.4cm}
\caption[]{\label{fig:Rthreshold} The spectral density
$R_{\tau}$ at low velocities in both QED and NRQED.}
\vspace*{-.5cm}
\end{figure}
\begin{figure}[!tb]
\includegraphics[angle=0,height=5.5cm,width=0.45\textwidth]{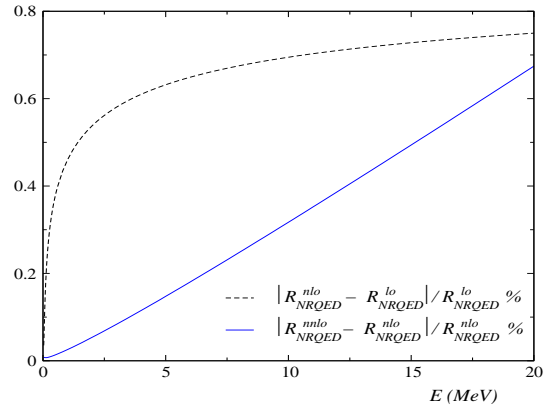}
\vspace*{-1.0cm}
\caption[]{\label{fig:Rsizes} Relative sizes of 
corrections to $R_{\tau}(s)$ as calculated in NRQED.}  
\vspace*{-.8cm}
\end{figure}
\begin{figure}[tb!]
\hspace*{-0.4cm}
\includegraphics[angle=0,height=6cm,width=0.5\textwidth]{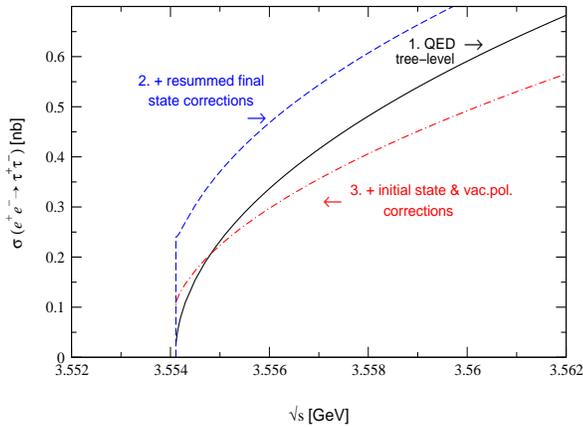}
\vspace*{-1.4cm}
\caption[]{\label{fig:kirk} Total cross section $\sigma(e^+e^-\to
\tau^+\tau^-)$ at threshold.}
\vspace*{-0.5cm}
\end{figure}

The need for performing resummations of the leading non-relativistic terms 
$\left( \alpha/v \right)^n [v ,v\alpha,v^2,\dots]$ is evidenced
in Fig.~\ref{fig:Rthreshold}. The QED spectral density vanishes as $v \to 0$,
due to
the phase space velocity in formula (\ref{s0}), which is canceled by the
first $v^{-1}$ term appearing in the ${\cal O}(\alpha)$ correction, making the
cross section at threshold finite. More singular terms near threshold,
$v^{-2},\dots$
arising in higher-order corrections completely spoil the expected good 
convergence of the QED perturbative series at low $v$. 
This is no longer the case for the effective theory
perturbative series, whose convergence improves as we approach the threshold
point, as shown in Fig.~\ref{fig:Rsizes}, and higher-order corrections
reduce the perturbative uncertainty. In the whole energy range displayed in Fig.~\ref{fig:Rsizes}, the
differences between the NNLO, NLO and LO results are
below 0.8\%, which indicates that the LO result, i.e. the
Sommerfeld factor, contains the relevant physics to describe the threshold
region.
Adding the intermediate and initial state corrections we have a complete
description of the total cross section of $\tau^+\tau^-$ production, as
shown in Fig.~\ref{fig:kirk}. Coulomb interaction between the produced $\tau$'s,
becomes essential within few
MeV's above the threshold, and its effects have to be taken into account to all
orders. 
Initial state radiation effectively reduces the available center-of-mass energy
for $\tau$ production, lowering in this way the total cross section. 

We should emphasize that NNLO corrections do not modify
the predicted behaviour of the LO and NLO cross section,
but are essential to
guarantee that the truncated perturbative series at NLO gets small
corrections from higher-order terms. Differences with previous evaluations
mainly concern the correct resummation of NNLO terms 
$\propto (\alpha/v)^n \times [\alpha^2,v^2,\alpha\,v]$, which on perturbative
grounds means that our results should differ by $\alpha^2$ terms. The latter
are seemingly negligible, but one has to check that all-order resummation of
them does not produce any unexpected enhancement. 

In addition, factorized 
formulas proposed in previous works do not have under control the 
NNLO terms 
being resummed beyond the order in $\alpha$ at which the matching between the 
non-relativistic and the QED results is performed. These fake resummations may
be harmless in QED, being $\alpha$ such a small parameter, but must be handled
with care when estimating the uncertainties of the theoretical
calculation. In that sense, notice in Fig.~\ref{fig:Rsizes}
that NNLO corrections to the spectral density are not 
${\cal{O}}(\alpha^2)\sim 0.005\%$ with
respect to the leading-order, as we would naively estimate.
Hence, we can safely conclude that the
theoretical uncertainty of our analysis of $\sigma(e^+e^- \to \tau^+\tau^-)$ 
at energies close to threshold is
lower than 0.1\%. 

A FORTRAN code which evaluates the spectral density 
$R^{\mbox{\tiny NNLO}}_{\mbox{\tiny $\tau$}}(s)$
will be soon available at {\bf http://alpha.ific.uv.es/$\sim$ruiz}

\bigskip \noindent
{\bf Acknowledgements}

I wish to thank Abe Seiden and his team for the organization
of the 7th Tau Workshop and for the opportunity to present this talk.
This work has been supported in part by the EU Research Training
Network EURIDICE No. HPRN-CT-2002-00311, by MCYT (Spain) 
under grant FPA2001-3031, and by ERDF
funds from the European Commission.


\end{document}